# Uncooled bolometer response of a low noise La$_{2/3}$Sr$_{1/3}$MnO$_3$ thin film

Laurence Méchin*, Jean-Marc Routoure, Bruno Guillet, Fan Yang, Stéphane Flament, Didier Robbes

GREYC (CNRS - UMR 6072), ENSICAEN and University of Caen, 6 boulevard du Maréchal Juin, 14050 Caen cedex, France

Radoslav A. Chakalov

School of Physics and Astronomy, University of Birmingham, Birmingham B15 2TT, UK

* Corresponding author (Email: lmechin@greyc.ensicaen.fr)

**Abstract**

We report measurements of the optical responses of a La$_{2/3}$Sr$_{1/3}$MnO$_3$ (LSMO) sample at a wavelength of 533 nm in the 300 – 400 K range. The 200 nm thick film was grown by pulsed laser deposition on (100) SrTiO$_3$ substrate and showed remarkably low noise. At 335 K the temperature coefficient of the resistance of a 100 µm wide 300 µm long LSMO line was 0.017 K$^{-1}$ and the normalized Hooge parameter was $9 \times 10^{-30}$ m$^3$, which is among the lowest reported values. We then measured an optical sensitivity at I = 5 mA of 10.4 V.W$^{-1}$ and corresponding noise equivalent power (NEP) values of $8.1 \times 10^{-10}$ W.Hz$^{-1/2}$ and $3.3 \times 10^{-10}$ W.Hz$^{-1/2}$ at 30 Hz and above 1 kHz, respectively. Simple considerations on bias current conditions and thermal conductance G are finally given for further sensitivity improvements using LSMO films. The performances were indeed demonstrated on bulk substrates with G of 10$^{-3}$ W.K$^{-1}$. One could expect a NEP reduction by three orders of magnitude if a membrane-type geometry was used, which makes this LSMO device competitive against commercially available uncooled bolometers.

The remarkable electronic and magnetic properties of rare-earth manganite oxides have raised lot of interests for applications [1]. In addition to the widely studied magnetoresistive sensors and spin electronics, uncooled bolometers using the large resistance change at the metal-to-insulator transition are among the most promising. Indeed uncooled technologies open new opportunities for infrared detection for both military and commercial applications [2]. Actual materials used for the fabrication of uncooled resistive bolometers are amorphous semiconductors, polycrystalline SiGe, semiconducting $YBa_2Cu_3O_7$ (YBCO), $VO_2$ and $VO_x$. Around 300 K they present typical maximum temperature coefficient of the resistance (TCR), defined as the relative resistance derivative $(1/R) \times (dR/dT)$, of 0.02 to 0.06 $K^{-1}$ [3, 4], -0.007 $K^{-1}$ [5], -0.029 $K^{-1}$ [6], -0.017 $K^{-1}$ [7, 8] and -0.033 $K^{-1}$ [9], respectively. Rajeswari et al. [10] showed that the optical response of $La_{0.67}Ca_{0.33}MnO_3$ was of thermal nature, but the maximum of the resistance derivative (corresponding to TCR = 0.079 $K^{-1}$) was obtained at 260 K. Goyal et al. [11] mentioned the high TCR values of several compositions of manganites. The highest values (0.2 $K^{-1}$) were obtained in $Nd_{0.7}Sr_{0.3}MnO_3$ well below room temperature (235 K). Only $La_{0.7}Sr_{0.3}MnO_3$ and $La_{0.7}Ba_{0.3}MnO_3$ that showed TCR of 0.02 $K^{-1}$ and 0.04 $K^{-1}$, respectively, were suitable for uncooled operation. Other manganite compositions such as $La_{2/3}(Ca,Pb)_{1/3}MnO_3$ [12] and $(La_{0.6}Pr_{0.4})_{0.67}Ca_{0.33}MnO_3$ [13] even showed TCR as high as 0.1 $K^{-1}$ at 300 K. However no noise data are available for the first material and preliminary measurements showing a high low frequency noise level are reported for the second one. In comparison TCR of $La_{0.5}Ba_{0.5}MnO_3$ and $La_{0.5}Sr_{0.5}MnO_3$ were lower (about 0.015-0.02 $K^{-1}$) but with a much lower noise level at 30 Hz, which makes them potentially more suitable for uncooled bolometer fabrication. Lisauskas et al. [14] tailored the TCR and the operating temperature by adding Pb to $La_{0.7}Sr_{0.3}MnO_3$. The resulting $La_{0.7}(Pb_{0.63}Sr_{0.37})_{0.3}MnO_3$ thin films exhibited TCR of 0.074 $K^{-1}$ at 295 K but a low frequency noise level about 5 orders of magnitude higher than has been reported in LSMO [15].

In addition to the remarkable resistance variation one has to consider the intrinsic noise in the thin films since it obviously sets the limit to the device performances. Continuous efforts have to be produced in order to understand the origin of noise and consequently to reduce the noise-to-signal ratio. Among manganites we concentrated our study on ferromagnetic $La_{1-x}Sr_xMnO_3$ with x ~ 0.3 (LSMO), which attracts especially great interests due to its high Curie temperature (~ 360 K), thus letting expecting room temperature devices, but also because it has shown so far remarkably low noise compared to other manganites [15, 16]. We therefore decided to optimize the bolometer performances by noise lowering rather than TCR increase. After a brief description of sample preparation and experimental conditions, this paper first gives the measured optical responses in the 300 – 400 K range. Noise measurements are presented, which enabled the calculation of the noise equivalent power (NEP) of the device. A comparison to NEP values of other uncooled bolometers is made and a discussion is finally given for further improvements.

The 200 nm thick LSMO thin film was deposited by pulsed laser deposition from a stoichiometric target onto (100) $SrTiO_3$ single crystal substrate. The laser radiation energy density was 1.5 J.cm$^{-2}$, the target-to-substrate distance was 65 mm, the oxygen pressure was 0.700 Torr and the substrate temperature was 780 °C. The X-ray diffraction study indicated a highly (100) orientation of the LSMO film. The resistivity of the film at 300 K was 26 μΩ.m. After LSMO deposition a 500 nm thick gold layer was sputtered on the films in order to make low resistive four-probe connections. The LSMO thin films were patterned by UV photolithography and argon ion etching for forming the 100 μm wide 300 μm long line. The contacts between the sample holder and the patterned film were finally made using aluminum wires by ultrasonic bonding. Optical measurements were performed using an Adlas DPY 315C frequency doubled semiconductor laser emitting a 533 nm beam, which was focused on the film surface. The diameter of the resulting laser spot was estimated to be 15 μm and the incoming power P was measured to be 3.6 mW (after the optical path). The sample was mounted on a X-Y micrometric table. For noise and optical response measurements, a four probe configuration was used. The dc voltage at the voltage probes of the sample was measured using a voltmeter (at 1 mA) simultaneously to the ac or "noise" voltage, thus enabling resistance and noise measurements, respectively. The noise level was measured using a commercial spectrum analyzer (HP3562A) at the output of a low-noise home-made amplifier of gain 10000 whose input impedance was adapted to the measured resistance. The voltage and current noise spectral densities of the amplifier were measured to be 1.5 nV.Hz$^{-1/2}$ and 3.2 pA.Hz$^{-1/2}$ at 10 Hz, respectively. In all measurements, the noise level was obtained by dividing the output noise by the gain of the amplification chain. No other corrections were made for all the noise signals.

Figure 1 shows the electrical resistance change $\Delta R = R_{on} - R_{off}$ when the laser spot position was varied along the X axis defined in inset. $R_{on}$ is the resistance of the bridge with the spot on it and $R_{off}$ is the resistance with no light. The resistance change is quite constant within the linewidth and drops sharply. The scan width is consistent with a linewidth of 100 µm, and a spot of about 15 µm in diameter. The change of the resistance $\Delta R$ with the laser spot in the middle of the bridge and incoming power P of 3.6 mW was recorded as a function of the temperature and compared with the derivative of the resistivity versus temperature (Fig. 2). Inset of figure 2 shows the very good superimposition of the curves normalized to unity at their maximum. It reveals that the laser does not generate charge carriers and that the optical response is purely of bolometric nature. The maximum value of the resistance derivative was obtained at 335 K and was equal to 12 Ω. The corresponding relative resistance derivative $(1/R) \times (dR/dT)$ was then equal to 0.017 K$^{-1}$, which is a typical value for the LSMO material with this composition [17, 18, 19]. We also checked that the resistance versus temperature characteristics did not vary with bias current in the 1 mA to 5 mA range, thus showing that there was no self heating effect. The corresponding maximum resistance change due to the spot at 335K was 7.5 Ω for a line resistance of 700 Ω. One can therefore deduce the measured optical voltage sensitivity $\Re_V$ of the line expressed in V.W$^{-1}$ as:

$$\Re_V = \frac{\Delta R \times I}{P} \quad (1)$$

where I is the bias current (A). For I = 2 mA and 5 mA, we thus obtained $\Re_V$ = 4.2 V.W$^{-1}$ and 10.4 V.W$^{-1}$ at 335 K, respectively. The definition of the optical voltage sensitivity of a bolometer pixel of resistance $R_0$ is usually calculated as [20]:

$$\Re_V = \frac{\eta \times I}{G_{eff} + jC\omega} \times \frac{dR_0}{dT} \quad (2)$$

where η is the absorption coefficient (dimensionless), C is the thermal capacitance (J.K$^{-1}$), $G_{eff}$ is the effective thermal conductance (W.K$^{-1}$), defined as $G_{eff} = G - I^2 \times (dR_0/dT)$ where G is the geometrical thermal conductance. $G_{eff}$ actually defines the conditions of thermal runaway for thermometers with positive resistance derivative in case of current biasing. One should be aware of the fact that equation (2) actually gives the sensitivity of a pixel of resistance $R_0$, constituting the total detecting area. In order to compare to our experiments that were made using a focused laser spot, we had to correct the resistance value by considering the spot as a pixel, and therefore considering the resistance of the area under illumination equal to the resistance-per-square $R_\square$. In such a way the total line resistance can be considered to be equal to 3×$R_\square$. We measured independently an absorption coefficient of 0.85. The ratio of the calculated dc sensitivity using equation 2 over the measured optical sensitivity (equation 1) at 335 K and 1 mA gives $G_{eff}$ = 1.6×10$^{-3}$ W.K$^{-1}$. This value is consistent with the thermal conductance of about 10$^{-3}$ W.K$^{-1}$ that has been measured for epitaxial YBCO films deposited on SrTiO$_3$ and that was attributed to the thermal boundary conductance at the film-substrate interface [21]. If G is roughly estimated to be equal to 10$^{-3}$ W.K$^{-1}$, the maximum bias current in order to avoid thermal runaway is 9 mA.

Figure 3 shows a typical example of the measured voltage noise density spectra at 300 K for several bias currents. The spectra clearly consist of two parts: a low frequency excess noise that depends on the bias current and the frequency, and the white noise that is current and frequency independent. The measured white noise level of $10^{-17}$ $V^2.Hz^{-1}$ corresponds well to the predicted Johnson-Nyquist value of 4kTR, with R = 700 Ω and where k is the Boltzmann constante and T is the temperature. The slope of the low frequency part was checked to be equal to -1 in all the investigated bias current and temperature ranges. The low frequency noise can therefore be considered as "1/f noise", which is usually described by the Hooge's empirical relation [22]:

$$\frac{S_V}{V^2} = \frac{\alpha_H}{n} \times \frac{1}{\Omega \times f} \qquad (3)$$

where $S_V$ is the voltage spectral density ($V^2.Hz^{-1}$), V is the applied voltage between the probes (V), $\alpha_H$ is the dimensionless Hooge parameter, n is the charge carrier density ($m^{-3}$), Ω the sample volume ($m^3$) and f the measuring frequency (Hz). Equation (3) is an empirical relation that does not have any physical statement. Its validity shows that fluctuations originate from resistance fluctuations. It has been verified in conventional metals and semiconductors, where typical values of $\alpha_H$ are about $10^{-3}$. $\alpha_H/n$ is defined as the normalized Hooge parameter and is expressed in $m^3$. It is very useful to use the normalized Hooge parameter in order to compare the magnitude of the 1/f noise in different materials independently of the volume and the bias conditions of the studied samples. We checked that $S_V$ varies quadratically with the applied voltage in all the investigated temperature range. We therefore calculated the normalized Hooge parameter $\alpha_H/n$ at each temperature (inset of figure 3). The measured value of $7.5 \times 10^{-30}$ $m^3$ at 300K is among the lowest reported for LSMO. Palanisami *et al.* [15] reported indeed $2.3 \times 10^{-32}$ $m^3$ on $SrTiO_3$ substrate but others obtained $\alpha_H/n$ values of $1.6 \times 10^{-26}$ $m^3$ on MgO substrate [17], $2\times10^{-28}$ m$^3$ on LaAlO$_3$ substrate, $10^{-26}$ - $10^{-24}$ m$^3$ on SrTiO$_3$ substrate [18] and $5.6\times10^{-24}$ m$^3$ on SrTiO$_3$ substrate [19].

The noise equivalent power (NEP) of the LSMO line can be estimated as the ratio of the measured voltage noise density (figure 3) over the bolometer voltage sensitivity:

$$NEP(f) = \frac{P \times \sqrt{S_V(f)}}{\Delta R \times I} \qquad (4)$$

Figure 4 gives its evolution as a function of the temperature at four different frequencies: 1 Hz, 10 Hz, 30 Hz and f > 1 kHz (the latter corresponding to the Johnson noise regime) and two bias currents (2 and 5 mA). The minimum measured NEP value was $3.3\times10^{-10}$ W.Hz$^{-1/2}$ at 330 K, I = 5 mA and f > 1 kHz. It was $8.1\times10^{-10}$ W.Hz$^{-1/2}$ at 30 Hz. For comparison, the NEP of semiconducting YBCO was measured to be $10^{-9}$ W.Hz$^{-1/2}$ at 30 Hz and at a wavelength of 9 μm using a Ti absorber (and $3.2\times10^{-10}$ W.Hz$^{-1/2}$ at 2 Hz) in case of 400 μm × 400 μm size membrane-type bolometers (G = $1.7\times10^{-5}$ W.K$^{-1}$) [6]. Minimum NEP values of $2\times10^{-11}$ W.Hz$^{-1/2}$ were obtained using amorphous semiconductors and optimized geometries [3]. The NEP measured by Chen *et al.* [8] in the 8-12 μm range at 30 Hz, 300 K and 20 μA was $1.67\times10^{-11}$ W.Hz$^{-1/2}$. They used a linear VO$_x$ microbolometer array fabricated on a micromachined substrate. In the case of manganites La$_{0.7}$(Pb$_{0.63}$Sr$_{0.37}$)$_{0.3}$MnO$_3$ thin film based bolometer showed a NEP of $3\times10^{-8}$ W.Hz$^{-1/2}$ at 30 Hz and 295 K despite a high TCR value of 0.074 K$^{-1}$ because of the higher 1/f noise level ($\alpha_H/n = 3\times10^{-27}$ m$^3$) [14].

Simple considerations can be made for the further NEP reduction of our LSMO bolometers. First the coefficient η could easily be increased by the deposition of an absorbing layer. Secondly equation (4) and figure 4 show that in the Johnson regime (f > 1 kHz), where the voltage noise density is equal to $\sqrt{4kTR}$, a possible optimization (NEP decrease) would be to increase the bias current. The maximum bias current was previously estimated to be 9 mA in order to avoid thermal runaway, thus leading to a minimum achievable optical NEP value of 1.86×10$^{-10}$ W.Hz$^{-1/2}$ at 330 K for this 100 μm wide 300 μm long line. In the low frequency regime, our measured NEP value does not depend on the bias current (figure 4) since it writes (after equation 3 and 4):

$$NEP(f) = \frac{P \times R}{\Delta R} \times \sqrt{\frac{\alpha_H}{n \times \Omega \times f}} \qquad (5)$$

Above mentioned NEP values extracted from literature were obtained on optimized membrane-type geometries. In contrast our NEP values were measured on bulk substrates without any geometry optimization. The thermal conductance G could largely be lowered by the use of silicon micromachining techniques, as was achieved with epitaxial YBCO films [23]. G values of about 10$^{-6}$ W.K$^{-1}$ could be obtained, thus resulting in NEP reduction by three orders of magnitude if we assume that such a low-noise film can be deposited on silicon substrates. In this paper we thus demonstrated the potential use of epitaxial LSMO thin films with TCR=0.017 K$^{-1}$ for uncooled bolometer fabrication. The relatively low TCR can largely be compensated by a lower noise compared to other manganite materials, vanadium oxides, amorphous and polycrystalline semiconductors, which make that LSMO bolometers can compete favorably with the commercially available uncooled bolometers.

**Acknowledgments**

The authors would like to acknowledge Prof. D.P. Almond who made the optical reponse measurement facility available for this study at the University of Bath (UK).

**References**

[1] T. Venkatesan, M. Rajaswari, Z. Dong, S.B. Ogale and R. Ramesh, Phil. Trans. R. Soc. Lond. A **356** 1661 (1998).

[2] J.L. Tissot, Infrared Phys. & Technol. **46** 147 (2004).

[3] V.Y. Zerov, V.G. Malyarov, J. Opt. Technol. **68** (12) 939 (2001).

[4] V.G. Malyarov, J. Opt. Technol. **69** (10) 750 (2002).

[5] S. Sedky, P. Fiorini, M. Caymax, C. Baert, L. Hermans, R. Mertens, IEEE Trans. Electron. Dev. **46** (4) 675 (1999).

[6] M. Almasri, Z. Çelik-Butler, D.P. Butler, A. Yaradanakul, A. Yildiz, J. Microelectromech. Syst. **11** (5) 528 (2002).

[7] A.P. Grudzeva, V.Y. Zerov, O.P. Konovalova, Y.V. Kulikov, V.G. Malyarov, I.A. Khrebtov, I.I. Shaganov, J. Opt. Technol. **64** (12) 1110 (1997).

[8] C.H. Chen, X.J. Yi, J. Zhang, X.R. Zhao, Infrared Phys. & Technol. **42** 87 (2001).

[9] V.Y. Zerov, V.G. Malyorov, I.A. Khrebtov, Y.V. Kulikov, I.I. Shaganov, A.D. Smirnov, J. Opt. Technol. **68** (6) 428 (2001).

[10] M. Rajeswari, C.H. Chen A. Goyal, C. Kwon, M.C. Robson, R. Ramesh, T. Venkatesan, S. Lakeou, Appl. Phys. Lett. **68** (25) 3555 (1996).

[11] A. Goyal, M. Rajeswari, R. Shreekala, S.E. Lofland, S.M. Bhagat, T. Boettcher, C. Kwon R. Ramesh, T. Venkatesan, Appl. Phys. Lett. **71** (17) 2535 (1997).

[12] Y. Sun, M.B. Salomon, S.H. Chun, J. Appl. Phys. **92** (6) 3235 (2002)

[13] R.J. Choudhary, A.S. Ogale, S.R. Shinde, S. Hullavarad, S.B. Ogale, T. Venkatesan, R.N. Bathe, S.I. Patil, R. Kumar, Appl. Phys. Lett. **84** (19) 3846 (2004).

[14] A. Lisauskas, S.I. Khartsev, A. Grishin, Appl. Phys. Lett. **77** (5) 756 (2000).

[15] A. Palanisami, R.D. Merithew, M.B. Weissman, M.P. Warusawithana, F.M. Hess, J.N. Eckstein, Phys. Rev. B **66** 092407 (2002).

[16] L. Méchin, J.-M. Routoure, S. Mercone, F. Yang, M. Saib, S. Flament, R.A. Chakalov, submitted to Phys. Rev. B

[17] B. Raquet, J.M.D. Coey, S. Wirth, S. von Molnar, Phys. Rev. B **59** 12435 (1999).

[18] A. Lisauskas, S.I. Khartsev, J. Low Temp. Phys. **117** Nos. 5/6 1647 (1999).

[19] K.-H. Han, Q. Huang, P.C. Ong, C.K. Ong, J. Phys.: Condens. Matter. **13** 8745 (2001).

[20] P.L. Richards, J. Appl. Phys. **76** 1 (1994).

[21] M. Nahum, S. Verghese, P.L. Richards, Appl. Phys. Lett. **59** 2034 (1991).

[22] F.N. Hooge, Phys. Lett. **29A** (3) 139 (1969).

[23] L. Méchin, J.C. Villégier, D. Bloyet, J. Appl. Phys. **81** (10) 7039 (1997).

**Figure captions**

Figure 1 (color online): Resistance variation $\Delta R = R_{on} - R_{off}$ of the LSMO line when the 15 µm diameter spot was scanned across the bridge in the X direction at T = 335 K where $R_{off}$ = 700 Ω. The incoming power was 3.6 mW.

Figure 2 (color online): Resistance versus temperature characteristic (left axis) of the 100 µm wide 300 µm long LSMO line, derivative of the resistance as a function of the temperature and resistance change (right axis) due to laser illumination (3.6 mW). Inset shows the dR/dT and ΔR curves normalized to unity at their maximum in order to show their very good superimposition.

Figure 3 (color online): Voltage noise density spectra of the 100 µm wide 300 µm long LSMO line for bias current in the 2 – 5 mA range. Inset shows the evolution of the normalized Hooge parameter $\alpha_H/n$ as a function of the temperature.

Figure 4 (color online): Measured NEP for bias currents of 2 mA and 5 mA as a function of the temperature (Note that closed and opened squares and circles are superimposed for f = 1 Hz and 10 Hz).

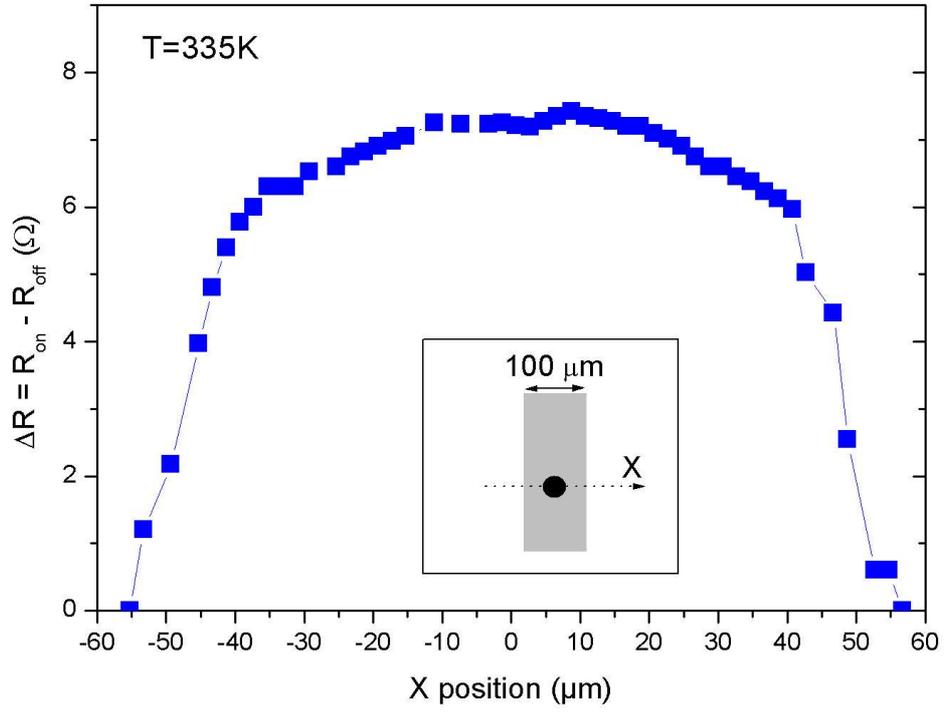

Figure 1 (color online)

Mechin *et al.*

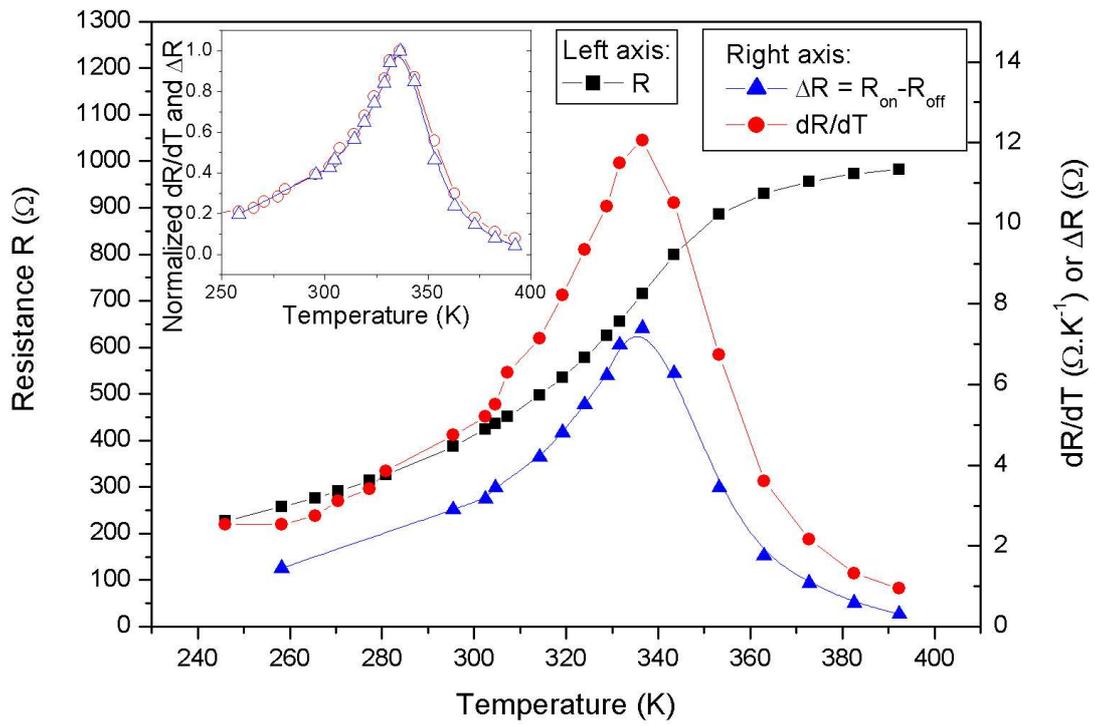

Figure 2 (color online)

Mechin *et al.*

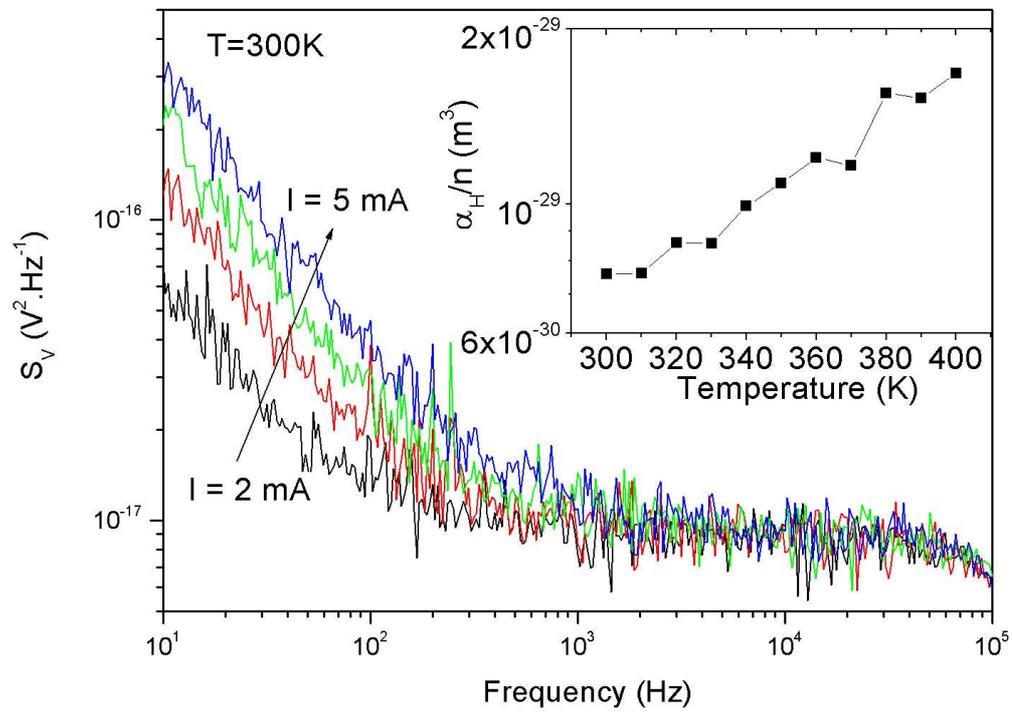

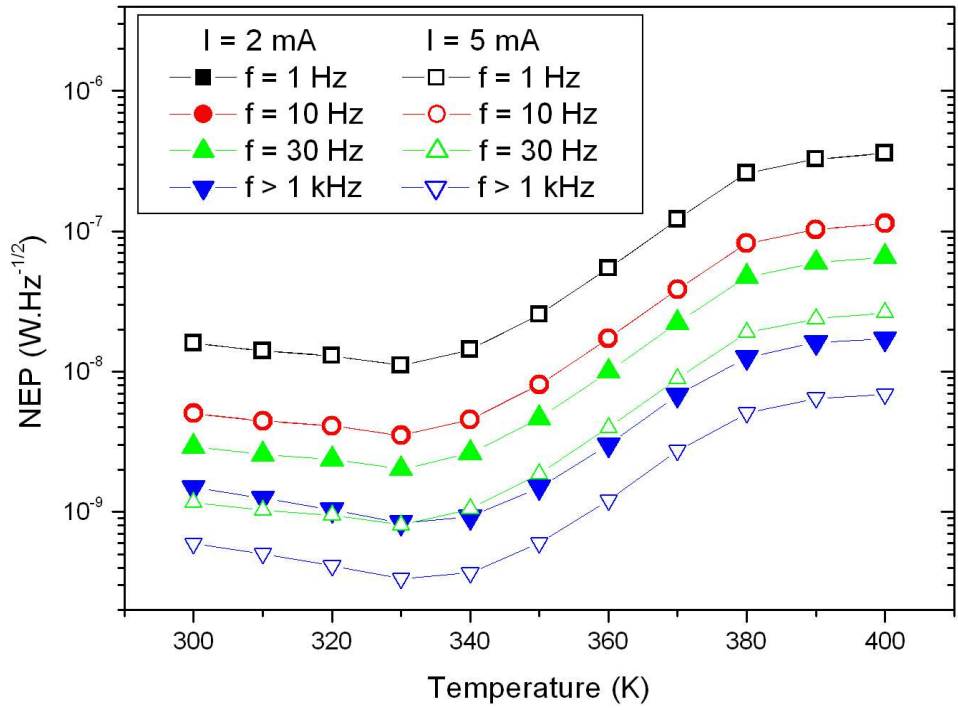

Figure 4 (color online)

Mechin *et al.*